\documentclass[12pt]{article}

\usepackage[a4paper,
            top=2.5cm, bottom=2.5cm,
            left=2.5cm, right=2.5cm]{geometry}

\usepackage[utf8]{inputenc}
\usepackage[T1]{fontenc}

\usepackage{amsmath,amssymb,amsfonts}
\usepackage{bm}          % bold math: \bm{n}

\usepackage{graphicx}
\usepackage{xcolor}

\usepackage{booktabs}

\usepackage{titlesec}
\titleformat{\section}{\large\bfseries}{\thesection}{1em}{}
\titleformat{\subsection}{\normalsize\bfseries}{\thesubsection}{1em}{}
\titleformat{\subsubsection}{\normalsize\itshape}{\thesubsubsection}{1em}{}

\usepackage{float}

\usepackage[labelfont=bf,font=small,labelsep=period]{caption}

\usepackage{setspace}
\usepackage[numbers,sort&compress]{natbib}
\usepackage[colorlinks=true,
            linkcolor=blue,
            citecolor=blue,
            urlcolor=blue]{hyperref}

\newcommand{\mum}{\,\mu\text{m}}

\newcommand{\Dx}{\Delta x}
\newcommand{\Dy}{\Delta y}

\begin{document}

% ---- title block -------------------------------------------
\begin{center}
  {\Large\bfseries
   Field-Programmable Topological Torons in Chiral Nematic Liquid Crystals\par}

  \vspace{1.4em}

  {\small
   Adithya Pradeep$^{a,\ast}$,
   Urban Mur$^{a,b}$,
   Ji Qin$^{a}$,
   Jonghyeon Ka$^{a,d}$,
   Waqas Kamal$^{a}$,\\[4pt]
   Tianxin Wang$^{a}$,
   Junseok Ma$^{a,c, d}$
   Jianming Wang$^{a}$,
   Steve~J.\ Elston$^{a,\ast}$,
   Stephen~M.\ Morris$^{a,\ast}$\par}

  \vspace{0.8em}

  {\small
   $^{a}$Department of Engineering Science, University of Oxford,
   Parks Road, Oxford OX1 3PJ, UK\\[2pt]
   $^{b}$Faculty of Mathematics and Physics, University of Ljubljana,
   Jadranska cesta 19, 1000 Ljubljana, Slovenia\\[2pt]
   $^{c}$Nature Sciences Research Institute, KAIST, 291 Daehak-ro,
   Yusong-gu, Daejeon 34141, Republic of Korea\\[2pt]
   $^{d}$Department of Electrical Engineering, POSTECH,
   Pohang 37673, Republic of Korea\par}

  \vspace{0.6em}

  {\small
   $^\ast$Correspondence:
   \href{mailto:adithya.nair@eng.ox.ac.uk}{adithya.nair@eng.ox.ac.uk},\;
   \href{mailto:steve.elston@eng.ox.ac.uk}{steve.elston@eng.ox.ac.uk},\;
   \href{mailto:stephen.morris@eng.ox.ac.uk}{stephen.morris@eng.ox.ac.uk}\par}

  \vspace{1em}
  \today
\end{center}

\vspace{1em}

% ---- abstract ----------------------------------------------
\begin{abstract}
\noindent
Torons are three-dimensional double-twist solitons in chiral nematic liquid crystals that form localised director configurations protected by topology and bounded by closed defect loops. They behave as particle-like entities while retaining a fully reconfigurable optical response. Here it is shown experimentally that individual torons can be created, steered and parked on demand using tailored alternating-current electric fields in planar cells, enabling deterministic control of both position and trajectory. By tuning the ratio of cell thickness to cholesteric pitch and systematically adjusting waveform parameters, including amplitude, modulation frequency, duty-cycle asymmetry and small DC offsets, robust toron nucleation is achieved and programmable translation is realised along arbitrary in-plane directions with submicrometre placement accuracy. Directional transport is controlled within a defined frequency and temperature window and can be reversed by changing modulation conditions even at zero offset. A dedicated graphical interface enables real-time switching between waveform presets so that torons follow scripted paths and draw user-defined shapes. Quantitative Landau–de Gennes Q-tensor simulations reproduce toron nucleation and the ensuing translational dynamics, supporting an interpretation in which waveform-controlled director reorientation, reorientation-driven flow and rectified polarity-sensitive coupling jointly bias the drift. Finally, three proof-of-concept functions are demonstrated: a software-defined liquid-crystal racetrack memory analogue with optical readout, deterministic path writing for reconfigurable patterning, and toron-mediated pick-and-place transport of microparticles for micromanipulation.
\end{abstract}

\vspace{0.5em}
\hrule
\vspace{1.5em}

% ============================================================
\section{Introduction}
% ============================================================

Liquid crystals (LCs) merge fluidity ~\cite{Leslie2016} with anisotropic molecular order ~\cite{Khoo2022}  and thus enable a diverse range of reconfigurable soft-matter architectures ~\cite{Bisoyi2021,Schwartz2018,%
Geryak2014,Jiang2019,Uchida2022}. By applying optical fields ~\cite{McConney2013,Ravnik2013}, electric fields ~\cite{Pradeep2026,Harkai2020} or patterned alignment ~\cite{Nys2020,Berteloot2020}, the LC director can be programmed into distinct topological configurations that remain stable under continuous external fields ~\cite{Tai2023}. These topological configurations enable a broad set of functionalities in optics, including spin-to-orbital angular momentum conversion with dynamic holography that imprints optical vortices ~\cite{Brasselet2011}, structured light emission and tunable microlasers ~\cite{Papic2021}, and reconfigurable photonic filters ~\cite{Isaacs2014} and smart windows ~\cite{Ma2025,ma2025advanced}. Beyond optics, topological textures and defects serve as elastic templates for colloidal assembly ~\cite{Wang2016} and directed particle transport ~\cite{ONeil2020} and as a temperature-sensitive label for transport ~\cite{Pradeep2025}.

Introducing molecular chirality into a nematic LC yields a chiral nematic (cholesteric) phase in which the helical pitch defines an intrinsic length scale and reshapes the elastic free-energy landscape ~\cite{Wu2022}. Within this frustrated landscape, three-dimensional solitonic textures known as torons can form ~\cite{Ackerman2012}: a local double-twist cylinder capped by closed disclination loops that confine the twist and suppress the associated splay and bend distortions.

A toron is a three-dimensional skyrmion-like soliton composed of a 
double-twist cylinder closed into a toroidal shape, with the twisted 
region terminated either by a pair of point defects of opposite hedgehog 
charge (radial $+1$ and hyperbolic $-1$) or by closed disclination loops, 
depending on boundary conditions~\cite{Smalyukh2010}. The double-twist 
structure rotates the director by approximately $180^{\circ}$ between the 
centre and the periphery of the toroidal cross-section, while the 
termination at the two point defects satisfies homeotropic surface 
anchoring at the confining plates~\cite{Selinger2016}. Torons can be 
regarded as localised ``atoms'' of twist embedded within an otherwise 
nearly uniform background~\cite{Ackerman2016}.

Traditionally, torons are generated in homeotropic cells by tightly focused, high-power laser writing of micrometre-scale structures ~\cite{Loussert2014}. In these geometries, the absence of a preferred in-plane axis leaves torons effectively stationary after formation, limiting applications that require controlled positioning. Here we generate torons in planar, antiparallel-rubbed cells, where the imposed director alignment introduces a well-defined in-plane directionality. Under applied alternating electric fields, this directionality enables reproducible nucleation, guided translation and precise placement of individual torons along user-programmed trajectories. Using electric fields rather than laser writing provides non-invasive and fully reversible control, allowing torons to be created, transported and erased on demand. Beyond functionality, our platform also offers a controllable route to interrogate how flexoelectric coupling ~\cite{Buka2013,Barnik1978} mediates the dynamics of topological solitons in liquid crystals, using torons as a model system.

In this work, we create these torons and understand its stability in a unform far field and elucidate electric-field-driven motion of torons and demonstrate controllable translation and placement in user-specified directions. We understand the stability of these torons by high speed imaging. Individual torons are steered along arbitrary in-plane trajectories and positioned at chosen locations with high precision. We quantify the dependence of motion on driving frequency and temperature, measure velocities, and perform trajectory and positional-error analyses to establish accuracy and repeatability. Representative functions are shown in the figures, including racetrack-memory-style prototype and programmable microparticle transport.

% ============================================================
\section{Results}
% ============================================================

\subsection{Creation of torons in liquid crystals}

\begin{figure}[H]
  \centering
  \includegraphics[width=\linewidth]{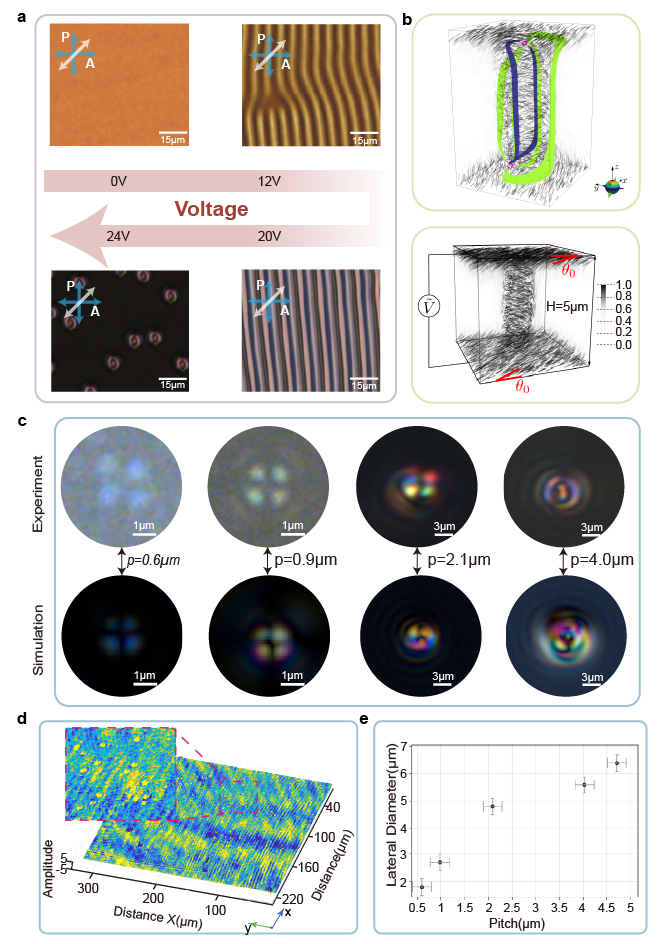}
  \captionsetup{justification=justified, width=\linewidth}
  \caption{\textbf{Voltage-induced formation and pitch-dependent scaling
    of torons in a confined chiral nematic cell.}
    (a)~POM micrographs at 0, 12, 20 and 24\,V showing the transition
    from a Grandjean stripe texture to isolated torons. Polariser (P) and
    analyser (A) directions indicated. Scale bars, 15\,$\mu$m.
    (b)~Numerical reconstruction of a toron: director field with coloured
    preimage surfaces (top) and simulation geometry, $H = 5\,\mu$m,
    under applied voltage (bottom).
    (c)~Experimental (top) and simulated (bottom) POM textures for varying
    pitch $p$. Scale bars, 1\,$\mu$m (left) and 3\,$\mu$m (right).
    (d)~DHM amplitude map showing the optical path-length modulation.
    (e)~Toron lateral diameter versus pitch.}
  \label{fig:creation}
\end{figure}

Torons were generated using electric fields in planar-aligned devices filled with a positive-dielectric chiral nematic mixture. The LC was confined between two antiparallel-rubbed substrates, imposing a uniform far-field director and a well-defined in-plane directionality that was later exploited for field-driven positioning. Unless stated otherwise, the chiral mixture comprised the nematic LC E7 (95.56\,wt\%) doped with the left-handed chiral additive S811 (5.00\,wt\%), with a helical pitch $p=2.062\mum$ in a cell of thickness $d=5\mum$ ($d/p=2.42$). Figure ~\ref{fig:creation} combines a representative voltage-driven unwinding sequence recorded for a longer-pitch sample ($p=4\mum$) with pitch-dependent optical signatures and scaling trends (panels c–e).

For the sample in Fig.~\ref{fig:creation}(a) ($p=4\mum$, $d=5\mum$ $d/p=1.25$), the director executes approximately $d/p\approx1.25$ full $2\pi$ rotations across the gap, producing a Grandjean texture with discrete domain walls. Under an applied AC voltage, the cholesteric initially retained a uniform helical state at 0\,V (Fig.~\ref{fig:creation}(a)). For a positive-dielectric-anisotropy mixture, the field favoured director alignment along the voltage axis, and the competition between dielectric torque and twist elasticity increased the effective pitch. With increasing voltage, the helix progressively unwound, and near 26\,V the LC transformed into a uniform homeotropic state. In the intermediate regime, the twist walls ~\cite{Kamien2001} that were formed around 12\,V collapsed; just below the unwinding threshold, at 4\,V,, this collapse created torons that emerged from this collapsing of the twist wall and persisted as localised double-twist structures. The device schematic is shown in the lower inset of Fig.~\ref{fig:creation}(b), with blue arrows indicating the antiparallel rubbing directions that impose planar alignment at the substrates.

Representative visualisation of a toron under an applied electric field is shown in Fig.~\ref{fig:creation}(b). The pitch window for toron formation was then established while keeping the cell gap fixed, such that the helical pitch provided the primary intrinsic length scale. To map the pitch window for toron formation, the cholesteric pitch was varied from $0.4\,\mu$m to $5.6\,\mu$m while maintaining a constant thickness of $d = 5\,\mu$m. For each pitch, the applied AC voltage was 
swept through the same qualitative sequence of textures: a Grandjean state, a uniform linear helix appearance, and then a regime in which the twist walls shrank and collapsed as the system approached the critical unwinding into a homeotropic state. Just before this critical voltage, torons were formed as localised double-twist objects that persisted over a finite voltage range just below the unwinding threshold. Across the dataset, toron formation was observed for pitches between $0.59\,\mu$m and $4.71\,\mu$m, corresponding to $d/p \approx 8.4$ down to $1.06$, whereas outside this range the voltage response did not yield stable torons. Having established the formation window, the toron lateral diameter was measured as a function of pitch, revealing a clear monotonic scaling: larger pitch produced larger lateral toron size at fixed thickness, consistent with pitch setting the characteristic in-plane extent of the double-twist region. Representative polarising optical microscopy (POM) images of experimentally realised torons together with $Q$-tensor simulations for matched pitches are shown in Fig.~\ref{fig:creation}(c), and the resulting pitch dependence of lateral  diameter is summarised in Fig.~\ref{fig:creation}(e).

Digital holographic microscopy (DHM) was used to quantify toron morphology. The system was configured as an off-axis Mach--Zehnder interferometer~\cite{Liu2019}: a small angular tilt between the reference and object beams at the recombination beamsplitter introduced a spatial carrier, and the resulting interference pattern was recorded as a hologram on a camera. The complex optical field was reconstructed using a single-FFT method followed by phase unwrapping~\cite{Yamaguchi2006}. Phase maps were referenced to the front-substrate plane, and amplitude maps were normalised to the background.

Figure~\ref{fig:creation}(d) shows the unwrapped phase of a single toron, revealing a ring-like optical path-length modulation consistent with an effective refractive-index contrast between the toron core and the surrounding host. The colour scale encodes the optical phase, while the corresponding amplitude map delineates the lateral extent of the toron. Voltage-dependent phase and amplitude reconstructions, including three-dimensional amplitude renderings versus lateral position, capture the full evolution of the cholesteric texture under the applied field. Raw holograms acquired at successive voltages spanning the transition from the Grandjean state to toron formation further confirm the robustness of the technique.

\subsection{Stability}

Torons remained stable under a uniform far field because stability is 
encoded both in their energetically favourable double-twist core and in 
the topological confinement provided by the accompanying defect loops. 
The double-twist structure locally accommodates chirality efficiently, 
reducing elastic penalties relative to twist distortions, while the 
defect loops bound the twisted region and prevent it from relaxing 
continuously into the surrounding background. In the present planar 
geometry, application of an electric field stabilises a homeotropic far 
field, and torons persist within this uniform background over a finite 
driving window. They were eliminated only when the field was driven 
outside this window, either restoring the uniform linear helix texture 
at lower drive or pushing the system towards complete unwinding at higher 
drive. For the mixture with pitch $p = 2.087\,\mu$m, twist walls began 
to contract at 8\,V (1\,kHz) and torons formed at 12\,V, before being 
destroyed at higher voltages as the system approached full unwinding.

\begin{figure}[H]
  \centering
  \includegraphics[width=\textwidth]{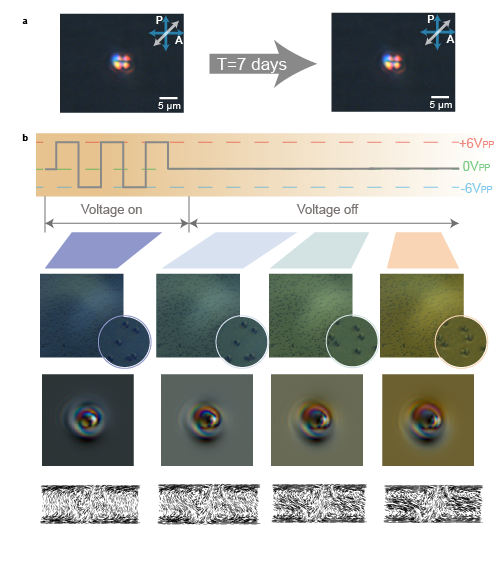}
  \caption{%
    \textbf{Field-stabilised persistence and relaxation dynamics of an
    isolated toron.}
    (a)~Polarising optical micrographs showing long-term stability under
    continuous driving, with no observable change after 7 days.
    (b)~Relaxation following field removal. Top: applied voltage protocol
    ($\pm6$\,V$_{\mathrm{pp}}$ square-wave drive followed by a zero-field
    interval). Middle: experimental POM time sequence showing the toron
    core persisting while the surrounding background destabilises. Bottom:
    corresponding $Q$-tensor simulations and director profiles reproducing
    the same relaxation pathway. Scale bars, 5\,$\mu$m.
  }
  \label{fig:stability}
\end{figure}

The temporal robustness of these solitons under sustained driving was 
assessed by nucleating torons at 12\,V and holding the sample at the 
same constant AC field for seven days. Polarising optical microscopy 
(POM) images recorded throughout this period showed no qualitative 
change in toron topology or optical signature, demonstrating long-term 
persistence under continuous external fields and stability within a 
uniform far field. Figure~\ref{fig:stability}(a) summarises this 
stability in planar cells when the applied drive is maintained below 
the unwinding threshold.

To probe stability under a deliberately changing far field, high-speed 
microscopy was used to resolve the step-by-step response of an individual 
toron to an abrupt modification of the drive. Because a toron is a 
topological soliton, its topology, including the number of holes in the 
texture, is preserved under continuous deformations provided the background 
boundary conditions remain intact. The voltage was therefore switched off 
rapidly from 12\,V to 0\,V while recording high-speed POM images. Notably, 
the toron did not disappear at the instant of switch-off. Instead, it 
remained intact initially, while the surrounding chiral nematic background 
began to reorganise away from the field-stabilised homeotropic state. This 
far-field evolution is evident as a progressive change in POM colour and 
texture outside the toron, reflecting a redistribution of the director 
field around the soliton (Fig.~\ref{fig:stability}(b)).

As this exterior director relaxation proceeds, the evolving far-field 
texture encroaches on the toron and progressively perturbs the boundary 
conditions that confine its double-twist core. Once the interference 
becomes sufficiently strong, topological protection can no longer be 
maintained within the changing background and the toron undergoes 
break-up. The high-speed POM sequence in Fig.~\ref{fig:stability}(b) 
captures this pathway directly: an initially unchanged toron immediately 
after switch-off, followed by eventual destruction driven by 
reorganisation of the region beyond the soliton.

Landau--de Gennes $Q$-tensor simulations performed under matching 
conditions corroborate this interpretation. The simulated POM images in 
the second column of Fig.~\ref{fig:stability}(b) reproduce the same 
sequence: following switch-off (12\,V to 0\,V), the toron persists while 
the surrounding texture evolves, visible as a colour change outside the 
toron that tracks reorientation of the external director field. As 
relaxation continues, the simulated far field invades the soliton and 
triggers its eventual destruction, consistent with the experimental 
observations. The corresponding director-field visualisations in 
Fig.~\ref{fig:stability}(b) further resolve the microscopic pathway, 
confirming that toron stability is maintained until the evolving exterior 
director field disrupts the confining topology and precipitates break-up.

\subsection{Movement}

Under a spatially uniform electric field, torons nucleated from 
twist-wall collapse remain essentially stationary. In conventional 
homeotropic cells this immobility is expected: the background is 
in-plane isotropic, with no preferred lateral axis to bias motion, 
so torons pin at sporadic locations set by local imperfections and 
nucleation history. Their positions are therefore random and 
irreproducible across the device, which limits applications that 
require deterministic placement, transport, or rewriting. By contrast, 
planar cells with antiparallel rubbing impose a well-defined in-plane 
director axis, providing the broken symmetry required for controlled, 
directional motion under electrical driving.

A first indication that the formation pathway itself can bias position 
was obtained by voltage cycling. After formation, a toron was tuned to 
a lower voltage at which extended twist walls reappear (10\,V, 1\,kHz). 
On reducing the drive, the localised toron elongated and relaxed back 
into an extended twist-wall segment; returning the voltage to 12\,V 
then regenerated the toron, demonstrating reversibility under repeated 
cycling. Notably, the re-formed toron did not return to its original 
lateral position. Instead, each down and up voltage cycle produced a 
finite net in-plane displacement, revealing a lateral step associated 
with the formation and collapse pathway. This reproducible net shift 
indicates an asymmetry in the collapse and re-nucleation dynamics and 
provides a practical mechanism to bias toron position under otherwise 
spatially uniform fields.

Continuous, deterministic translation was driven by applying a temporally 
modulated AC waveform to the planar cell. The drive consisted of a 1\,kHz 
sawtooth carrier whose peak amplitude was controlled by a 60\,Hz sawtooth 
envelope that ramped from 0\,V to 24\,V and then reset at the start of 
the next cycle. This repeated sweep traversed the toron stability window 
and generated a net in-plane bias: individual torons translated 
reproducibly in a single, fixed direction across the field of view and 
could be repositioned along user-defined paths. Motion persisted under 
sustained modulation and ceased promptly when the envelope modulation was 
halted, even though the 1\,kHz carrier remained applied. The response was 
repeatable across multiple runs with identical parameters, and trajectories 
were consistent for torons initialised at different starting locations 
within the same device.

A custom control interface was developed to apply prescribed voltages 
and waveforms during toron manipulation. The software functioned as a 
waveform synthesiser, enabling real-time specification of amplitude and 
offset, selection of waveform shape (including continuous interpolation 
between sawtooth and square forms), control of waveform asymmetry, and 
independent tuning of the carrier frequency and the low-frequency 
envelope modulation. Toron motion was monitored in situ by polarising 
optical microscopy using a CCD camera, and videos were recorded for 
subsequent trajectory and velocity analysis.

Under burst-modulated driving, torons exhibited a pronounced directional 
drift (northward in the laboratory frame for the waveform shown). As the 
modulation frequency was increased, the drift speed decreased and vanished 
at 72\,Hz, where the toron became effectively stationary. Increasing the 
drive amplitude at fixed frequency then induced a reversal of the drift 
direction, and the magnitude of the reversed velocity increased with 
amplitude, persisting up to 90\,Hz. Beyond 90\,Hz, no net translation 
was detected and stable torons could not be maintained under otherwise 
identical conditions. These trends were reproducible across repeated 
frequency sweeps on the same device.

Because torons emerge from the collapse of twist walls, their field 
response is expected to inherit key features of twist-wall 
electrodynamics. Twist walls exhibit a flexoelectric contribution under 
temporally asymmetric driving that mimics a chiral flexo-electro-optic 
response governed by $e_1 - e_3$~\cite{Elston2008}. To impose an 
analogous bias on torons, temporal symmetry was deliberately broken in 
the applied waveform. Starting from a 1\,kHz carrier with 60\,Hz 
amplitude modulation and a peak of 24\,V, a symmetric sawtooth (equal 
rise and fall, zero DC offset) produced translation perpendicular to the 
rubbing axis (north in the laboratory frame). The trajectory angle was 
then rotated by introducing waveform skew: with a rise-time fraction of 
20\% and a fall-time fraction of 80\% (still at zero DC offset), torons 
translated obliquely towards the rubbing direction (north-west), 
demonstrating that duty-cycle asymmetry alone biases the angle of motion. 
Increasing the asymmetry further to a 99\% rise-time fraction and adding 
a small DC offset of $+200$\,mV generated translation parallel to the 
rubbing axis towards the west. Reversing the sawtooth skew or changing 
the sign of the DC offset inverted the drift towards the east. Together, 
rise and fall asymmetry and a small DC bias provided a compact control 
space for selecting the in-plane translation angle over the full 
$0$--$180^{\circ}$ range permitted by the planar alignment, while speed 
and persistence remained constrained by the modulation-frequency window 
identified above. The response was reproducible over repeated runs, 
insensitive to starting position within a device, and reversible on 
demand by switching between waveform presets. Piecewise switching of 
asymmetry and offset enabled scripted trajectories including straight 
segments parallel or perpendicular to rubbing, oblique diagonals, and 
closed paths; motion ceased promptly when the envelope modulation was 
halted, even if the 1\,kHz carrier remained applied. The dependence on 
waveform polarity and temporal asymmetry is consistent with a directional 
bias arising from the interplay between a time-asymmetric drive and the 
broken in-plane symmetry imposed by planar alignment, without requiring 
a specific microscopic mechanism.

\begin{figure}[H]
  \centering
  \includegraphics[width=\textwidth]{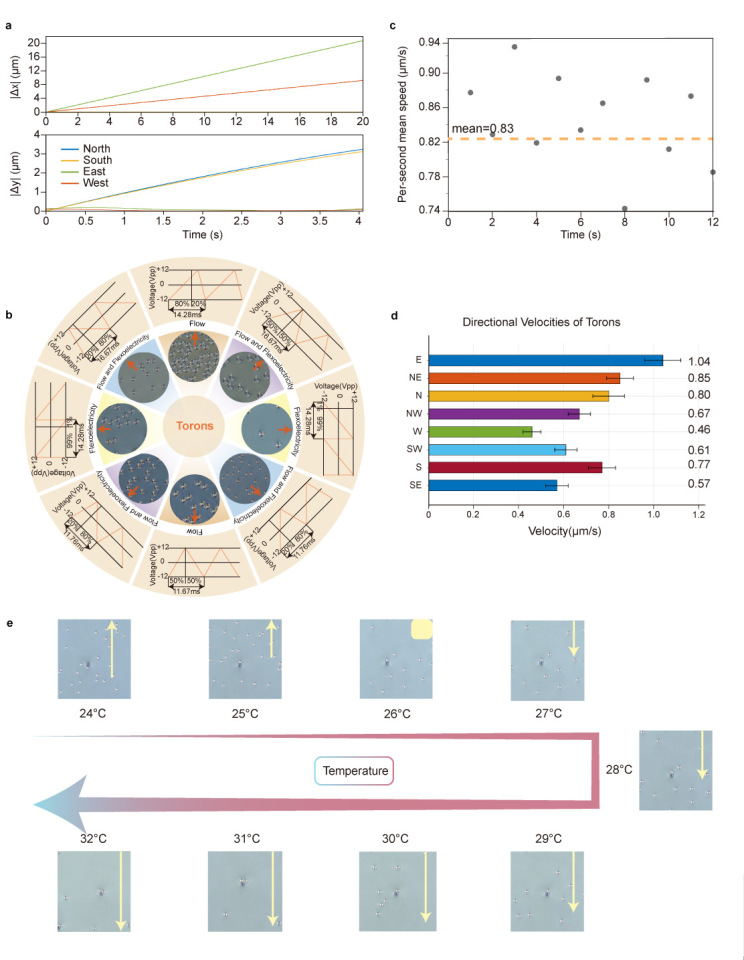}
  \caption{%
    \textbf{Programmable directional transport of torons using
    burst-modulated electric fields and temperature-controlled reversal.}
    (a)~Position versus time for a representative toron driven along the
    four cardinal directions ($\Dx(t)$ top, $\Dy(t)$ bottom).
    (b)~Waveform library enabling eight-direction control.
    (c)~Per-second mean speed over a 12\,s interval for the north-east
    trajectory; dashed line indicates the trimmed mean.
    (d)~Mean velocities for all eight programmed directions. Error bars
    denote the standard error of per-second means from 12 one-second bins.
    (e)~Temperature-controlled reversal of transport under a fixed
    electrical drive; images acquired in 1\,$^{\circ}$C steps. All
    measurements used a 1\,kHz carrier; ``north'' is defined perpendicular
    to the rubbing direction.
  }
  \label{fig:transport}
\end{figure}

Position control was quantified by plotting $\Delta x(t)$ and 
$\Delta y(t)$ for representative motion along the cardinal directions 
(north, south, east and west). The waveforms used were: north, a 
symmetric sawtooth (50\% rise-time fraction) with a 60\,Hz modulation 
frequency and a peak amplitude of 24\,V; south, a symmetric sawtooth 
(50\%) at 85\,Hz with a peak amplitude of 24\,V; east, a strongly 
asymmetric sawtooth (99\% rise-time fraction) at 70\,Hz with a peak 
amplitude of 24\,V and a $+200$\,mV DC offset; and west, an asymmetric 
sawtooth (1\% rise-time fraction) at 70\,Hz with a peak amplitude of 
24\,V. The resulting $\Delta x(t)$ and $\Delta y(t)$ traces were well 
approximated by linear trends, indicating near-constant speed over the 
observation window. A waveform library enabling motion in eight 
programmed directions is summarised in Fig.~\ref{fig:transport}(b).

Trajectories were extracted from microscopy videos using a 
pixel-to-micrometre conversion set by a user-defined reference in the 
first frame, based on a single toron of known diameter. Bright features 
corresponding to torons were segmented frame by frame, small spurious 
regions were removed, and detections near the top and bottom image 
margins were excluded. Centroids were linked between frames using 
nearest-neighbour association with a maximum permitted step to avoid 
identity swaps, and displacements were reported using a fixed sign 
convention such that motion towards image north is positive. For visual 
clarity, low-order polynomial guides were overlaid on $\Delta x(t)$ and 
$\Delta y(t)$, whereas velocities were obtained quantitatively from 
linear fits to the position--time segments.

Velocity analysis was used to quantify toron motility under 
burst-modulated driving. Instantaneous speeds were computed from 
successive centroid positions and converted to $\mu$m\,s$^{-1}$. 
Outliers were rejected if the implied single-frame jump exceeded 
20\,$\mu$m or if the instantaneous speed exceeded 10\,$\mu$m\,s$^{-1}$. 
Per-second mean speed was then obtained by averaging valid framewise 
speeds within each 1\,s bin, and a trimmed mean was reported after 
discarding the upper 5\% of per-second values. A representative 
mean-speed trace over a 12\,s window is shown in 
Fig.~\ref{fig:transport}(c). Figure~\ref{fig:transport}(b) further 
demonstrates on-demand steering in all eight directions through switching 
between waveform presets.

To distinguish motion dominated by flow from motion dominated by 
flexoelectric bias, mean velocities were compared across the programmed 
directions. Directions chosen to emphasise flow-driven transport yielded 
similar speeds, for example $0.80\,\mu$m\,s$^{-1}$ for northward motion 
at 60\,Hz and $0.77\,\mu$m\,s$^{-1}$ for southward motion at 85\,Hz. 
By contrast, directions designed to accentuate flexoelectric bias 
exhibited a broader spread of speeds, spanning both the highest and 
lowest velocities observed across the set. This wider dynamic range 
indicates that flexoelectric contributions provide a stronger and more 
tunable handle on toron velocity than flow alone, although both 
mechanisms may act concurrently depending on the waveform and operating 
point.

\subsection{Temperature effects}

All waveform-driven transport experiments described above were performed 
at a single room temperature, isolating how waveform parameters alone 
bias toron motion. Temperature dependence was then examined using the 
complementary protocol of holding the electrical drive fixed while 
varying temperature. Using a symmetric sawtooth waveform (1\,kHz 
carrier, 60\,Hz modulation, 24\,V peak), torons translated north at 
22\,$^{\circ}$C. The temperature was then increased from 24\,$^{\circ}$C 
to 34\,$^{\circ}$C in 1\,$^{\circ}$C steps with a 10\,min dwell at each 
setpoint. Between 24\,$^{\circ}$C and 25\,$^{\circ}$C, northward motion 
persisted but slowed markedly. At 26\,$^{\circ}$C, net translation 
vanished and the toron remained effectively stationary. On further 
heating to 27\,$^{\circ}$C, the drift reversed and torons moved south, 
with speed increasing monotonically with temperature thereafter.

In parallel, the density of observable torons decreased with increasing 
temperature, and no torons could be sustained at or above 33\,$^{\circ}$C 
under these driving conditions. This temperature-driven reversal echoes 
the frequency-driven reversal described above, indicating that temperature 
shifts the same operational window that governs directional transport.

Figure~\ref{fig:transport}(e) captures the same reversal when the sample 
is heated in 1\,$^{\circ}$C increments with a 10\,min dwell at each 
setpoint. Varying the heating rate did not measurably alter the direction 
of transport: the drift direction at a given temperature and fixed waveform 
was the same for ramps of 1\,$^{\circ}$C per 10\,min and 1\,$^{\circ}$C 
per 1\,min. This indicates that the reversal is set primarily by the 
instantaneous temperature under the constant 60\,Hz modulation protocol, 
rather than by the rate of temperature change.

\subsection{Applications}

This work establishes fully programmable creation, translation and 
erasure of individual torons in planar liquid-crystal devices, enabling 
deterministic positioning with submicrometre placement accuracy. 
Directional control across eight cardinal and diagonal directions, 
combined with GUI-scripted trajectory following, provides a general 
route to on-demand manipulation of stable topological solitons. In this 
framework, rise--fall (duty-cycle) asymmetry and small DC offsets bias 
the in-plane drift direction, while the modulation frequency sets the 
transport window and tunes the speed; notably, even with zero DC offset, 
changing the modulation frequency can reverse the drift direction (north 
to south).

These capabilities enable several practical application pathways. First, 
shuttling single torons between prescribed sites naturally realises 
racetrack-memory-style operation~\cite{Tomasello2014}, where a soliton 
serves as a rewritable, mobile information carrier within a planar 
architecture. Second, deterministic placement of localised double-twist 
regions provides a reconfigurable optical building block: toron arrays 
could be written, translated and erased to create dynamically addressable 
phase and scattering landscapes for beam shaping~\cite{Shealy2006}, 
filtering or adaptive micro-optics~\cite{Shealy2006,Zappe2015}. Third, 
toron transport offers a programmable micro-actuation mechanism for 
soft-matter micromanipulation, enabling controlled capture, transport and 
release of microparticles templated by the toron's elastic and topological 
fields. Together, electrically programmable toron positioning provides a 
practical platform for deploying stable topological textures in 
liquid-crystal technologies.

For many application scenarios, sparse populations are preferred so that 
individual torons can be addressed and routed independently. To access 
this regime, a low-density distribution of torons was generated by 
increasing the peak amplitude above 23\,V, which reduced the nucleation 
yield in the planar cell. A single toron was then selected and driven 
under GUI control. By switching between waveform presets that set the 
in-plane direction through duty-cycle asymmetry and a small DC offset, 
while maintaining a 1\,kHz carrier and 60\,Hz modulation, the toron was 
programmed to trace the letters ``SMP''. The trajectory remained 
continuous, with accurate placement at corners and reliable turning at 
preset switch points, illustrating repeatable point-to-point positioning 
and robust path following.

\begin{figure}[H]
  \centering
  \includegraphics[width=\textwidth]{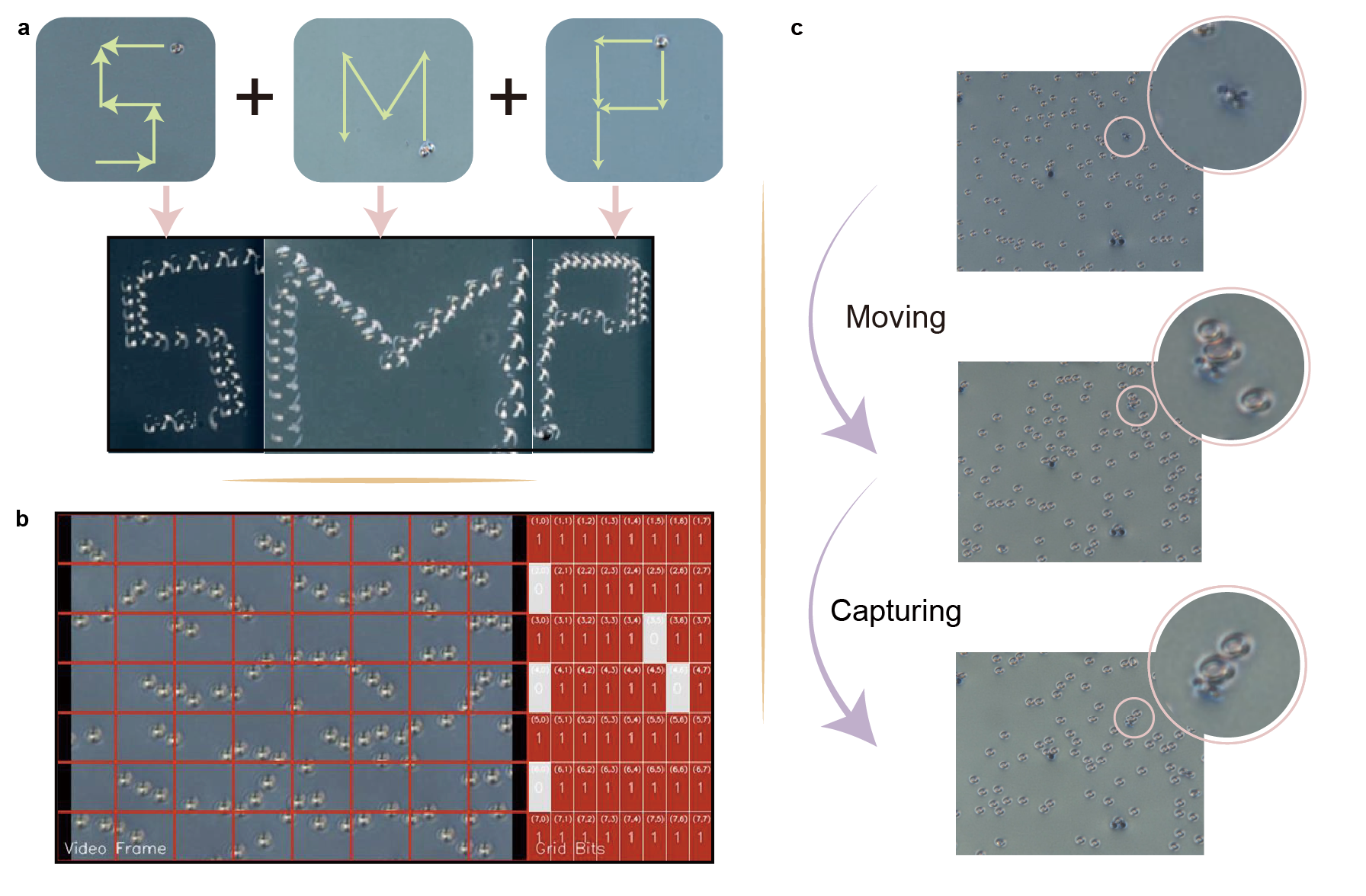}
  \caption{%
    \textbf{Proof-of-concept applications.}
    (a)~Centroid-tracked trajectory of a single toron steered to trace the
    letters ``SMP'' under GUI-controlled waveform presets.
    (b)~Liquid-crystal racetrack memory analogue: torons written, shifted
    and erased along a reconfigurable virtual track with optical readout.
    (c)~Toron-mediated microparticle pick-and-place: a silica bead cluster
    captured by the toron elastic field is transported and released at a
    target site. Scale bars, 10\,$\mu$m.
  }
  \label{fig:applications}
\end{figure}

Figure~\ref{fig:applications}(a) shows the centroid-tracked trajectory 
of a single toron steered using the GUI-controlled waveform library. The 
reconstructed path traces the letters ``SMP'', confirming on-demand 
control of both position and heading. Trajectories were extracted from 
bright-field videos by frame-by-frame centroid linking, and the overlaid 
track highlights faithful execution of straight segments and sharp turns, 
demonstrating repeatable point-to-point placement and continuous rendering 
of user-defined shapes.

A liquid-crystal racetrack memory analogue was realised by steering 
individual torons along a software-defined grid, where the occupancy of 
each site encodes a binary state that can be written, shifted and erased 
deterministically (Fig.~\ref{fig:applications}(b)). In contrast to 
magnetic racetrack concepts~\cite{Tomasello2014} that rely on 
lithographically defined nanowires, engineered pinning sites and 
current-driven spin torques, the track here is virtual and reconfigurable. 
Torons are routed using low-voltage AC waveforms, with waveform parameters 
tuned to execute straight translations and $90^{\circ}$ turns on demand 
through a combined contribution of director-reorientation backflow and 
flexoelectric bias. Bits are read optically from polarised-microscopy 
contrast and can be verified in real time by straightforward image 
segmentation.

This approach offers several practical advantages. Tracks can be 
reprogrammed in software without physical patterning, enabling serpentine 
paths, branch points and parallel lanes within the same device. The 
information carriers are topological solitons with long retention under 
constant bias, supporting low standby power. Operation is achieved at 
room temperature in planar ITO cells using modest amplitudes at kilohertz 
carrier frequencies, avoiding Joule heating and materials fatigue 
associated with current-driven spin devices. Together, optical readout 
combined with electrical write and shift, and on-the-fly reconfigurability, 
establishes a soft-matter racetrack platform with simple fabrication, high 
endurance and a natural route to dense two-dimensional arrays.

Microparticle transport was demonstrated by dispersing spacer beads in 
the cholesteric host and steering torons electrically as described above. 
The particles were nominally neutral, silica-based spacers with a diameter 
of 0.9\,$\mu$m, which agglomerated in the LC into clusters with an 
apparent lateral size of approximately 4\,$\mu$m. Under burst-modulated 
driving, freely moving clusters were captured by nearby torons and remained 
bound during subsequent translation. Once loaded, a toron acted as a mobile 
carrier, transporting the cargo along user-selected directions set by 
waveform presets and releasing it at chosen locations. 
Figure~\ref{fig:applications}(c) summarises the corresponding trajectories, 
highlighting pick-and-place operation and directional routing for 
microparticle distribution.

\section{Discussion}

This work establishes an electrically driven platform for the formation, 
guidance and precise positioning of torons in planar liquid-crystal cells. 
Torons are generated without structured optical beams or high-power laser 
and remain stable under sustained driving within a defined operating window. 
Landau--de Gennes $Q$-tensor simulations reproduce key elements of the 
experimental behaviour, including toron nucleation and the observed 
morphologies under field, providing independent validation of the identified 
textures and the field--director coupling that underpins their response.

Toron creation and stability are central to the transport mechanism. In 
planar antiparallel rubbed cells, the applied field drives a reproducible 
pathway in which twist walls collapse to nucleate localised double-twist 
torons that persist over a finite operating window. This controlled 
nucleation and retention under sustained drive provide a reliable starting 
state for subsequent waveform-based steering.

Mean translation speeds were measured under modulated drives for motion 
along eight programmed directions: north (N), south (S), east (E), west 
(W), and the diagonals NE, NW, SE and SW. Here, ``north'' is defined as 
motion perpendicular to the rubbing axis under a symmetric sawtooth 
waveform at 60\,Hz modulation with a 1\,kHz carrier. The observed trends 
are consistent with two waveform-tunable contributions to the drift. 
First, for a sawtooth drive with zero DC offset, any polarity-sensitive 
cycle-averaged bias is minimised; translation is then dominated by 
periodic director reorientation and the associated reorientation-driven 
flow, yielding motion primarily along the rubbing normal (N/S). Second, 
introducing a strongly asymmetric duty cycle together with a small DC 
offset rectifies the response and produces a non-zero cycle-averaged 
lateral bias along the easy axis, favouring E/W motion. Diagonal 
trajectories arise when both components are present, with the drift 
direction following the vector sum of a reorientation-driven north--south 
component and a rectified east--west component, and the relative weights 
set by waveform asymmetry and offset.

Accordingly, the drive recipes were selected to weight these contributions 
in a controlled manner. The ``Flow Up/Down'' settings employ a 
time-balanced sawtooth (50\% rise, 50\% fall) with zero DC offset at 60 
or 85\,Hz to emphasise reorientation-driven flow, which biases motion 
along the rubbing normal. The ``Flexo Right/Left'' settings use extreme 
duty-cycle asymmetry (1\% or 99\% rise fraction) together with a small 
DC offset to maximise polarity-sensitive coupling and thereby bias motion 
along the rubbing axis. The four diagonal modes adopt intermediate 
duty-cycle asymmetries with zero or very small offset, producing mixed 
contributions that steer NE, NW, SE or SW through the vector sum of the 
two components.

The velocity ordering in Fig.~\ref{fig:transport}(d), with E fastest and 
W slower than N/S, is consistent with waveform conditions that generate a 
larger rectified lateral component in the ``Flexo Right/Left'' modes than 
the impulse generated by the time-balanced waveform, while also 
highlighting that the relative weighting need not be symmetric between 
opposite polarities. Figure~\ref{fig:transport}(b) summarises this 
directional control by listing the waveform parameters used to obtain each 
direction, including the 1\,kHz carrier, modulation rate, peak voltage, 
duty-cycle asymmetry and any DC offset, and by indicating the dominant 
contribution emphasised by each recipe as defined above.

A key advance is directional programmability in a planar geometry. Using 
a 1\,kHz carrier with low-frequency amplitude modulation, and tuning 
duty-cycle asymmetry together with small DC offsets, torons can be steered 
along any of the eight cardinal and diagonal directions with submicrometre 
placement accuracy. A GUI enables rapid switching between waveform presets, 
supporting straight translations, $90^{\circ}$ turns, diagonals and 
scripted trajectories, including continuous drawing of letter shapes. These 
capabilities translate directly into two proof-of-concept functions. First, 
a liquid-crystal racetrack analogue is demonstrated in which occupancy on 
a software-defined grid is written, shifted and erased electrically, with 
optical readout and reconfigurable tracks that require no lithographic 
patterning. Second, torons act as micromanipulation carriers that capture, 
transport and release microparticles along user-selected routes, enabling 
soft-matter pick-and-place at the micrometre scale. Both functions operate 
at room temperature using modest voltages in standard planar ITO devices.

The approach provides a practical and tunable operating envelope for 
electrically steerable torons. Across a wide range of frequencies and 
temperatures, directional transport remains robust, with the unwinding 
transition acting as a natural boundary that defines the accessible 
control space. Within this envelope, waveform parameters and material 
properties jointly set speed, persistence and heading, offering multiple 
levers including mixture composition, anchoring strength and cell geometry 
to boost throughput, lower drive requirements and extend scalability to 
large arrays.In a wider photonics context, such deterministic control over localised soft-matter states may ultimately be combined with laser-written photonic architectures, where coupling gaps and waveguide termination geometry play an important role in device performance \cite{qin2025termination,qin2026gap}.

Several important questions are naturally opened by this platform. 
Crosstalk between neighbouring torons, many-body interactions and 
long-term endurance under repeated cycling remain to be quantified, and 
the control loop could be strengthened by closed-loop feedback using 
real-time imaging and automated correction of drift and placement error. 
Integration with patterned microelectrodes or active-matrix backplanes 
could enable dense two-dimensional addressability, supporting 
reconfigurable photonic elements, scalable memory architectures and 
multiplexed microrobotic transport. More broadly, the same waveform-based 
control framework should be transferable to other liquid-crystal 
topological textures, including skyrmions and hopfions, and to hybrid 
schemes that combine electrical steering with optical writing or readout.

In summary, electrically programmable creation, translation and erasure 
of individual torons in planar cells provides a general method for 
creation and positioning topological solitons on demand. The combination 
of simple fabrication, optical readout, low-voltage operation and software 
reconfigurability offers a practical route to liquid-crystal 
implementations of racetrack-style information processing, tunable 
photonics and micromanipulation.

\section{Methods}

\subsection{Liquid Crystal Mixture and Cell Preparation}

A left-handed cholesteric mixture was prepared by doping the nematic host 
E7 with the chiral additive S811 (Merck). The dopant level was selected 
to achieve the target helical pitch for the experiments. The mixture was 
introduced by capillary action into glass cells of 5\,$\mu$m gap, defined 
by spacer beads, at a temperature above the clearing temperature of the 
mixture. Both substrates carried indium tin oxide (ITO) electrodes to 
permit application of an electric field across the liquid crystal. Over 
the ITO, polyimide alignment layers were deposited and mechanically rubbed 
in antiparallel directions, yielding an antiparallel-rubbed cell. 
Electrical leads were attached to the ITO pads and connected to a waveform 
generator. The completed cell was mounted on the optical platform for 
observation and electrical driving. No reactive mesogen or photoinitiator 
was used in this study, and no polymer network was formed.

\subsection{Velocity Analysis}

Video analysis was performed in MATLAB R2024b (MathWorks) with the Image 
Processing Toolbox. Movies were sampled at 60\,Hz, yielding 
$n_F = \lfloor T \cdot 60 \rfloor$ analysed frames for a video of 
duration $T$ seconds.

Spatial calibration was carried out on the first frame. A circle tightly 
circumscribing a single toron was drawn, and the pixel-to-length scale $s$ 
was computed from the known toron diameter $D$ as $s = D/(2R)$, where $R$ 
is the drawn radius in pixels. All coordinates were scaled by $s$ to 
convert to micrometres.

Each frame was converted to grayscale and binarized by adaptive 
thresholding with bright foreground polarity and sensitivity 0.45. 
Connected components smaller than 10 pixels were removed, and binary 
holes were filled. Object centroids were obtained from region properties. 
To limit edge artifacts, detections with vertical coordinate $y \leq 40$ 
px or $y \geq H - 40$ px, where $H$ is the image height, were excluded.

Frame-to-frame linking used a greedy nearest-neighbour association. For 
each active track propagated from frame $f-1$, the candidate centroid in 
frame $f$ with the smallest Euclidean distance was accepted if the 
separation was less than 15\,px. Unassigned detections initiated new 
tracks, and tracks without an assignment in the next frame were terminated.

\subsubsection*{Mean Speed Analysis Calculation}

Instantaneous speed for interval $[f-1, f]$ was computed from successive 
positions $\mathbf{r}_f = (x_f, y_f)$ in micrometres:

\begin{equation}
    v_f = \frac{|\mathbf{r}_f - \mathbf{r}_{f-1}|}{\Delta t} = 
    |\mathbf{r}_f - \mathbf{r}_{f-1}| \times 60\,\mu\text{m}\,\text{s}^{-1}
\end{equation}

Intervals lacking a valid association were treated as missing values.

Temporal aggregation was performed at 1\,Hz by grouping consecutive sets 
of 60 frames into one-second bins. For bin $k$, the per-second mean speed 
$\bar{v}_k$ was defined as the mean of all valid instantaneous speeds 
within that bin. Where indicated for visualization only, values smaller 
than $0.2\,\mu$m\,s$^{-1}$ were set to NaN to suppress quiescent 
intervals.

Summary statistics were reported as a trimmed mean baseline $\hat{v}$, 
obtained by removing the upper 5 percent of $\{\bar{v}_k\}$ and averaging 
the remainder. This approach reduces sensitivity to rare mismatches during 
linking without relying on fixed speed thresholds.

Plots were produced in MATLAB R2024b (MathWorks) as scatter displays of 
$\bar{v}_k$ versus time $t_k = k$ s with large filled markers. Axes used 
outward ticks, the top and right spines were hidden, and a rectangular 
frame was drawn. Unless stated otherwise, fixed parameters were: maximal 
link distance 15\,px, excluded margins 40\,px, minimum object area 
10\,px, trimmed-mean percentile 5 percent, and optional visualization clip 
at $0.2\,\mu$m\,s$^{-1}$.

\subsection{Single-Toron Displacement Tracking and Polynomial Fitting}

Single-toron displacement analysis proceeded by selecting, after linking, 
the trajectory whose earliest centroid lay closest to the calibration 
location in the first frame. With positions recorded in micrometres as 
$(x_i, y_i)$ over frames $i = 1, \ldots, n$, the time axis was defined 
by the acquisition rate as $t_i = (i-1)/60$\,s. Displacements were 
referenced to the initial position and plotted with a fixed sign 
convention so that motion toward image north appears positive, namely

\begin{equation}
    \Delta x_i = -(x_i - x_1), \quad \Delta y_i = -(y_i - y_1)
\end{equation}

Intervals where the linker failed to associate a detection were left as 
gaps; the corresponding samples were marked missing and no interpolation 
across gaps was applied prior to fitting or plotting.

For presentation, each component was fitted independently with a low-order 
polynomial in time. Unless stated otherwise a quadratic model was used for 
both $\Delta x(t)$ and $\Delta y(t)$. The coefficients were obtained by 
least squares on the available $(t_i, \Delta x_i)$ and $(t_i, \Delta y_i)$ 
pairs. When the number of samples was insufficient for the requested 
degree, the raw displacement trace was retained without fitting. To convey 
the underlying measurements alongside the smooth trend, a set of evenly 
spaced time points was overlaid as circular markers; by default six markers 
were used per panel, distributed uniformly across the trace duration.

Figures displayed $\Delta x(t)$ and $\Delta y(t)$ in separate axes using 
box framing with outward ticks, Arial font, and grid lines disabled. 
Legends and titles were omitted in final layouts to preserve visual 
economy. Parameters held fixed across datasets, unless specified, were: 
frame rate 60\,Hz for the time base, excluded margins 40\,px and minimum 
object area 10\,px during detection, nearest-neighbour cap 15\,px for 
association, polynomial degree 2 for both components, and six evenly 
spaced sample markers.

\subsection{$Q$-tensor simulations}

We perform numerical simulations of the director field using the
phenomenological Landau--de Gennes (LdG) free-energy relaxation
method~\cite{Ravnik2009}, including the liquid-crystal order parameter,
elastic anisotropy, rotational viscosity, and dielectric coupling to the
external electric field.

The equilibrium configuration of the director field is obtained by
minimisation of the free-energy density functional, expressed in terms of
the liquid-crystal order-parameter tensor
\begin{equation}
  Q_{ij} = \tfrac{1}{2}S\!\left(3n_i n_j - \delta_{ij}\right),
\end{equation}
where $S$ is the scalar order parameter and $\mathbf{n}$ is the director.
The free-energy density contains multiple contributions. The phase
(Landau-expansion) part is
\begin{equation}
  f_{\mathrm{phase}} =
    \tfrac{1}{2}K\!\left(\partial_k Q_{ij}\right)^2
    -\tfrac{1}{3}\varepsilon_0\varepsilon_a^{\mathrm{mol}}\,Q_{ij}E_i E_j
    +\tfrac{1}{2}A(T)\,\mathrm{Tr}(Q^2)
    +\tfrac{1}{3}B\,\mathrm{Tr}(Q^3)
    +\tfrac{1}{4}C\!\left[\mathrm{Tr}(Q^2)\right]^2,
\end{equation}
where the coefficient $A = a(T - T_{\mathrm{NI}}^{*})$, $T$ is the
temperature, $T_{\mathrm{NI}}^{*}$ is the supercooling temperature, and
$a$, $B$, $C$ are constant material parameters. The elastic
free-energy contributions are
\begin{align}
  f_{\mathrm{el}} &=
    \tfrac{1}{2}L_1
      \frac{\partial Q_{ij}}{\partial x_k}
      \frac{\partial Q_{ij}}{\partial x_k}
    +\tfrac{1}{2}L_2
      \frac{\partial Q_{ij}}{\partial x_j}
      \frac{\partial Q_{ik}}{\partial x_k}
    +\tfrac{1}{2}L_3
      \frac{\partial Q_{kl}}{\partial x_i}
      \frac{\partial Q_{kl}}{\partial x_j}, \\[4pt]
  f_{\mathrm{chiral}} &=
    Lq_0\,\varepsilon_{ikl}\,Q_{ij}
    \frac{\partial Q_{lj}}{\partial x_k}, \\[4pt]
  f_{\mathrm{diel}} &=
    -\tfrac{1}{2}\varepsilon_0\bar{\varepsilon}
    -\tfrac{1}{3}\varepsilon_0\varepsilon_a^{\mathrm{mol}}\,Q_{ij}E_i E_j,
\end{align}
where $\partial_k$ denotes the partial derivative with respect to the
$k$-th spatial coordinate. The first term of $f_{\mathrm{el}}$ represents
the elastic free energy in a one-constant approximation, with $K$ the
Frank elastic constant under the isotropic elasticity assumption. The
dielectric coupling constant $\varepsilon_a^{\mathrm{mol}}$ enters both
the phase and dielectric free-energy terms.

Boundary conditions were fixed to
$\mathbf{n} = (\cos 85^{\circ},\, 0,\, \sin 85^{\circ})$ at the top and
bottom surfaces, corresponding to the experimental antiparallel-rubbing
geometry. The free energy was minimised using the Euler--Lagrange equation
\begin{equation}
  \frac{\partial f}{\partial Q_{ij}}
  - \partial_k \frac{\partial f}{\partial(\partial_k Q_{ij})} = 0.
\end{equation}

\subsection{Polarised Optical Microscopy}

Polarised microscopy imaging was carried out on an Olympus BX51 fitted 
with a QImaging Retiga R6 camera. Olympus objectives were used, and for 
each objective the cover-glass correction collar was tuned to the measured 
substrate thickness of the LC cell to minimise aberrations. A 550\,nm 
long-pass filter was inserted in the illumination path to block actinic 
light and prevent unintended polymerisation of any residual reactive 
mesogen. During acquisition the polariser and analyser were kept crossed. 
The device was rotated until the rubbing direction lay at $45^{\circ}$ to 
the polariser transmission axis, yielding a bright POM state; this 
orientation was kept fixed for all datasets.

Simulated POM images were generated using Nemaktis~\cite{Nemaktis}, an 
open-source Beam Propagation Method (BPM) software. Numerically computed 
director fields, described in the previous section, served as the input, 
and a broadband illumination spectrum was modelled by sampling 25 different 
wavelengths.

\subsection{High-Speed Imaging}

To characterize the temporal response and dynamic modulation behaviour 
of the toron when switched off, a high-time-resolution transmission 
microscope was constructed. A linearly polarized, collimated laser beam 
at a wavelength of 634.9\,nm (THORLABS CPS635) was used as the 
illumination source.

The imaging system consisted of a $50\times$ objective lens with a 
numerical aperture of 0.50 (OLYMPUS LMPlanFI) and a tube lens with a 
focal length of 150\,mm. In this calibration configuration, the LCCM 
was mounted on a precision translation stage and positioned at the sample 
(object) plane of the microscope, where it was directly imaged onto the 
camera. This arrangement allowed the spatiotemporal evolution of the 
speckle patterns generated by the LCCM to be measured directly. It is 
emphasized that this configuration was used solely for device 
characterization and is distinct from the illumination-modulation geometry 
employed in the imaging experiments described elsewhere in this work.

The modulated optical field was recorded using a high-speed camera 
(PHOTRON FASTCAM NOVA S6), operated at a frame rate of 20,000 frames per 
second with a spatial resolution of $512 \times 512$ pixels. Based on 
calibration with a USAF resolution target (THORLABS Negative 1951 USAF 
Test Target), the field of view of the imaging system was determined to 
be $108.1 \times 108.1\,\mu$m$^2$. Under these conditions, the microscope 
provided sufficient spatial resolution to resolve speckle patterns with 
characteristic sizes larger than $2\,\mu$m and sufficient temporal 
resolution to capture speckle dynamics on timescales down to 0.05\,ms.
% ============================================================
% References
% ============================================================
\bibliography{refs}

\end{document}